\documentclass[12pt,a4paper]{article}
\usepackage{type1cm,amsmath,hangcaption,indentfirst}
\usepackage[psamsfonts]{amssymb}
\numberwithin{equation}{section}

\newcommand{\br}{\mathbb R}
\newcommand{\bz}{\mathbb Z}
\newcommand{\bc}{\mathbb C}

\newcommand{\ket}[1]{{|#1\rangle}{}}

\newcommand{\nn}{\nonumber\\}
\newcommand{\ts}[1]{{\textstyle #1}}

\DeclareMathOperator*{\re}{{\rm Re}}
\DeclareMathOperator*{\im}{{\rm Im}}

\begin{document}
\begin{titlepage}
 
 \renewcommand{\thefootnote}{\fnsymbol{footnote}}
\begin{flushright}
 \begin{tabular}{l}
 SNUST-040901\\
 hep-th/0409185\\
 \end{tabular}
\end{flushright}

 \vfill
 \begin{center}
 \font\titlerm=cmr10 scaled\magstep4
 \font\titlei=cmmi10 scaled\magstep4
 \font\titleis=cmmi7 scaled\magstep4
 \centerline{\titlerm Boundary states in the Nappi-Witten model} 
 \vskip 2.5 truecm

\noindent{ \large Yasuaki Hikida}\footnote{
E-mail: hikida@phya.snu.phys.ac.kr}
\bigskip

 \vskip .6 truecm
\centerline{\it School of Physics \& BK-21 Physics Division,
                Seoul National University}
\vspace*{0.25cm}
\centerline{\it Seoul 151-747, Korea}
 \vskip .4 truecm

 \end{center}

 \vfill
\vskip 0.5 truecm

\begin{abstract}

We investigate D-branes in the Nappi-Witten model.
Classically symmetric D-branes are classified by the (twisted) 
conjugacy classes of the Nappi-Witten group, 
which specify the geometry of the corresponding D-branes.
Quantum description of the D-branes is given by boundary states,
and we need one point functions of closed strings to construct
the boundary states.
We compute the one point functions solving conformal bootstrap 
constraints, and check that the classical limit of the boundary states 
reproduces the geometry of D-branes.

\end{abstract}
\vfill
\vskip 0.5 truecm

\setcounter{footnote}{0}
\renewcommand{\thefootnote}{\arabic{footnote}}
\end{titlepage}

\newpage

\section{Introduction}
\label{Intoduction}

String theory on backgrounds of pp-wave type attracts much 
attention recently.
Geometry of pp-wave type appears in a (pp-wave) limit of
Anti-de Sitter (AdS) space and strings on pp-wave background
may be solvable. Many works on the string theory have been done
after the authors \cite{BMN} applied this fact to the AdS/CFT 
correspondence.
The spacetime of $AdS_5 \times S^5$ with RR-flux reduces to 
the maximally supersymmetric pp-wave \cite{mspp-wave1} 
by the pp-wave limit,
and the string theory on the pp-wave background is solvable in the 
light-cone gauge \cite{pp-wave1,pp-wave2} even with 
non-trivial RR-flux.
In Ref.~\cite{BMN} they compared almost BPS operators in
${\cal N} = 4$ super Yang-Mills theory and closed strings
on the pp-wave background.
D-branes in this background have been also investigated by many authors,
for example, in Refs.~\cite{ppdbrane2,KNS,ST1,BMZ,Hikida,ST2,ST3}.
In particular, the boundary states are constructed in 
Refs.~\cite{ppdbrane1,Green1,Green2,Green3} using the light-cone gauge.

The Nappi-Witten model \cite{NW} is a Wess-Zumino-Witten (WZW) model 
associated with 4 dimensional Heisenberg group 
$H_4$, whose target space is 4 dimensional pp-wave with NSNS-flux.
The pp-wave limit of $AdS_3 \times S^3$ with NSNS-flux is 6 dimensional
generalization of the Nappi-Witten model, and we can also apply
this model to the AdS/CFT correspondence
\cite{KP,HS1,LM,GMS,HS2,Narain,Hikidaphd}.
Because the Nappi-Witten model is a WZW model, we can do more than in
the case of the pp-waves with RR-flux.
The model can be solved without taking the light-cone gauge, 
and the correlation functions are obtained 
in Refs.~\cite{DK,CFS,BDKZ} recently.
The model itself is also very interesting apart from the application 
to the AdS/CFT correspondence since it is an example
which can be solved and has non-trivial {\it Lorentzian} target space-time.
In many cases non-trivial Lorentzian theory is defined by 
analytic continuation of Euclidean theory which may be solvable.
In the Nappi-Witten model, however, we can solve the model directly with
the Lorentzian signature, and there is no difficulty associated with
analytic continuation.

In this paper we investigate D-branes in the Nappi-Witten model.
For a non-rational conformal field theory, it is very difficult
to solve the theory generally, and in particular, 
boundary states are constructed only in few examples.
Because the Nappi-Witten model is a solvable non-rational conformal
field theory, it is worthwhile to construct boundary states also
for the respect.
We assume that the D-branes preserve the half of the symmetry of 
the current algebra, then the D-branes are classified by the 
(twisted) conjugacy classes \cite{NWDbrane1,NWDbrane2,NWDbrane3}.%
\footnote{For other types of D-branes in the pp-wave with
NSNS-flux see 
Refs.~\cite{Takayanagi,Michishita,Panigrahi1,Panigrahi2}.}
After reviewing the geometry of the target space-time and 
the closed strings of the model in section 2,
we examine the classical geometry of possible D-branes in section 3.
In section 4 we compute disk one point functions of closed
strings and construct boundary states for the D-branes.\footnote{%
Recently the one point functions are also investigated 
in Ref.~\cite{Hpp}.}
Section 5 is devoted to the conclusion.

\section{The Nappi-Witten Model}

The Nappi-Witten model \cite{NW} is a WZW model based on
4 dimensional Heisenberg group $H_4$. 
The generators of $H_4$ Lie algebra have the following non-trivial
commutation relations as
\begin{align}
 [J,P^{\pm}]&=\mp i P^{\pm} ~, &[P^+,P^-]&=-2iF ~.
 \label{CR}
\end{align}
A convenient way to parametrize the group element is
\begin{align}
 g(x^+,x^-,y) = e^{\frac12 x^+ J} 
                  e^{\frac{i}{\sqrt2 } (y^* P^- + y P^+)} 
                    e^{\frac12 x^+J-2x^-F}~,
\end{align}
where the group product is given by
\begin{align}
 &g(x^+_1,x^-_1,y_1) g(x^+_2,x^-_2,y_2) 
   \nn &\qquad = g(x^+_1 + x^+_2 , x^-_1 + x^-_2 
       +{\textstyle \frac{1}{2}} 
        \im (y_1 y^*_2 e^{\frac{i}{2}(x^+_1 + x_2^+)}) ,
        y_1 e^{\frac{i}{2} x_2^+}+ w_2e^{-\frac{i}{2} x^+_1}) ~.
         \label{glaw2}
\end{align}
The group element leads to the metric of pp-wave type as
\begin{equation}
 d s^2 = - 2 d x^+ d x^- - \frac{1}{4} y y^* (d x^+)^2 +
 d y d y^*~.
 \label{metric2}
\end{equation}
Here we use the definition
$ds^2 = \frac12 \langle g^{-1}dg , g^{-1}dg \rangle $ with
$\langle J , F \rangle = 1 , \langle P^+ , P^- \rangle = 2$.

The symmetry $g \to  g^*_L g g_R$ on the metric 
is generated by the following differential operators as%
\footnote{For later convenience we flip the
sign $P_L^{\pm} \to - P^{\pm}_L$.}
\begin{align}
 J_L &= \partial_{x^+} - {\textstyle \frac{i}{2}}
        (y \partial_y - y^* \partial_{y^*}) ~, 
& F_L &= - {\textstyle \frac12 } \partial_{x^-} ~,\nn
P^+_L &= - i \sqrt2 e^{-i\frac{x^+}{2}} 
  (\partial_{y^*} + {\textstyle \frac{i}{4}} y \partial_{x^-}) ~,
&P^-_L &= - i \sqrt2 e^{i\frac{x^+}{2}} 
  (\partial_y - {\textstyle \frac{i}{4}} y^* \partial_{x^-}) ~,
  \label{diffop1}
\end{align}
and the right part with replacing $y \leftrightarrow y^*$.
These operators are read from the  group multiplication law \eqref{glaw2}.

In the WZW model the symmetry of $H_4$ Lie algebra is enhanced 
to current algebra,
whose generators have the operator product expansions (OPEs)
\begin{align}
 J(z) F(w) &\sim \frac{1}{(z-w)^2} ~, \nonumber\\
 J(z) P^{\pm} (w) &\sim \frac{\mp i P^{\pm}(w)}{z-w} ~,
 \label{H4OPE} \\
 P^+(z) P^- (w) &\sim \frac{2}{(z-w)^2} - \frac{2iF(w)}{z-w} ~.\nonumber
\end{align}
The mode expansions of the currents satisfy
\begin{align}
  [ J_m , F_n ] & = m \delta_{m+n,0} ~,
& [ J_m , P^{\pm}_n ] & = \mp i P^{\pm}_{m+n} ~,
& [ P^+_m , P^-_n ] & = 2 m \delta_{m+n,0} - 2iF_{m+n} ~,
\label{H4CR}
\end{align}
where the zero-mode subalgebra is $H_4$ Lie algebra.
Anti-holomorphic (right-moving) currents are given in the similar way.

\subsection{Primary fields}

The general states in the WZW model can be constructed  by
\begin{align}
 &J^{A_1}_{- n_1} \cdots J^{A_l}_{- n_l} \ket{v} ~,
 &J^{A}_{n} \ket{v} &= 0 \quad ({}^{\forall} n > 0 ) ~,
\end{align}
and tensoring the anti-holomorphic part.
We use $n_i > 0$ and $A_i = (J,F,\pm)$ with $J^{J}= J$, $J^{F}=F$,
$J^{\pm} = P^{\pm}$.
Note that there is no singular vector in general. 
The state $\ket{v}$ is a vacuum state labeled by the representation of 
the zero-mode subalgebra, whose irreducible unitary representations 
are summarized e.g. in Refs.~\cite{KK,KKL}.

The vacuum state can be labeled by two eigenvalues as
\begin{align}
 J_0 \ket{j,\eta} &= i j \ket{j,\eta} ~,
&F_0 \ket{j,\eta} &= i \eta \ket{j,\eta} ~,
\end{align}
and $P^+_0$ and $P^-_0$ act on the eigenstates as the lowering and rising 
operators, respectively (see \eqref{CR}). 
As in Refs.~\cite{KK,KKL}, 
there are the following Hilbert spaces based on
three types of unitary representation.
One of them includes a lowest weight state $P^+_0 \ket{j,\eta} = 0$ 
and the other vacuum states are given by the action of $P^-_0$. 
The Hilbert space based on this representation is called as
${\cal H}_{j,\eta}^+$ with $j\in\br$ and $0<\eta<1$.
We will construct the states with $\eta > 1$ by applying
the spectral flow in the next subsection.
There is a similar Hilbert space ${\cal H}_{j,\eta}^-$ including the 
highest weight state $P^-_0 \ket{j,\eta} = 0$, and the general vacuum 
states are generated by acting $P^+_0$ to the highest weight state.
The labels of ${\cal H}_{j,\eta}^-$ range $j \in \br$ and $-1<\eta<0$.
The conformal weights for these vacuum states are
\begin{align}
h^{\pm} = \frac{|\eta|}{2}(1-|\eta|)-j\eta ~.
\end{align}
The other Hilbert space ${\cal H}^0_{j,s}$ does not include lowest nor 
highest weight state.
The general vacuum states are generated by acting $P^{\pm}_0$ to a
vacuum state with $-1/2 < j \leq 1/2$. 
The other parameter $s \,(\geq 0)$ is related to the conformal weight as
\begin{align}
h^0=\frac{s^2}{2} ~.
\end{align}
The eigenvalue of $F_0$ is zero ($\eta = 0$).

In order to express the primary fields corresponding to the vacuum states,
it is convenient to introduce the
parameter $x \,(\in \bc)$ to sum up the representation
of zero-mode subalgebra. 
In this parametrization we can map the action of the currents to the
primary fields into the differential operation as
\begin{align}
J^{A} (z) \Psi (w,x) \sim \frac{{\cal D}_x^{A} \Psi (w,x)}{z-w} ~.
\label{diffeq}
\end{align}
Therefore, the constraint of the symmetry may be written in the form of
differential equations, and in Ref.~\cite{DK} the correlation functions
are computed heavily using this property. 
The differential operators are \cite{DK}
\begin{align}
{\cal D}^J_x &= i(j+x\partial_x) ~,
&{\cal D}^F_x &= i \eta ~, 
&{\cal D}^+_x &= \sqrt2 \eta x ~,
&{\cal D}^-_x &= \sqrt2 \partial_x ~,
& (\eta > 0) \\
{\cal D}^J_x &= i(j - x\partial_x) ~,
&{\cal D}^F_x &= i \eta ~,
&{\cal D}^+_x &= \sqrt2  \partial_x~,
&{\cal D}^-_x &= - \sqrt2  \eta x~.
& (\eta < 0)
\end{align}
For $\eta = 0$ we use a phase $x=e^{i\alpha}$ and 
\begin{align}
{\cal D}^J_x &= ij + \partial_{\alpha} ~,
&{\cal D}^F_x &= 0 ~,
&{\cal D}^+_x &= s e^{i\alpha}~,
&{\cal D}^-_x &= s e^{-i\alpha}~.
\end{align}

Since the zero-mode of the currents correspond to 
\eqref{diffop1}, the classical expression
of the primary fields are obtained by solving the differential
equations \eqref{diffeq}. For ${\cal H}_{j,\eta}^{+}$
the solutions are\footnote{We construct the closed string spectrum using
the same representation in the holomorphic and anti-holomorphic part.}
\begin{align}
 \Psi^{+}_{j,\eta} 
   &= {\cal N}^+_{j,\eta} 
      \exp \left( i j x^+ - 2 i \eta x^- - \frac{\eta y y^*}{2} 
       + i \eta \left( 
         y \bar x e^{ i \frac{ x^+}{2} } + y^* x e^{ i \frac{ x^+}{2} } \right)
         + \eta x \bar x e^{ix^+} \right) \nn
   &= {\cal N}^+_{j,\eta} e^{ ijx^+ -  2 i \eta x^- - \frac{\eta y y^*}{2} }
      \sum_{m,n} \frac{\eta^m  y^{m-  n}}{m!}(i \bar x)^m (-i x)^{n} 
       e^{i \frac{ (m + n)x^+}{2}}  
       L^{m - n}_{ n} (\eta y y^*) ~,\label{wf+}
\end{align}
where $L^k_n (x)$ is the associated Laguerre polynomial.
Similarly for ${\cal H}_{j,\eta}^{-}$
 \begin{align}
 \Psi^{-}_{j,\eta} 
   &= {\cal N}^-_{j,\eta}
     \exp \left( i j x^+ - 2 i \eta x^- + \frac{\eta y y^*}{2} 
       + i \eta \left( 
         y x e^{- i \frac{x^+}{2} } + y^* \bar x e^{- i \frac{x^+}{2} } 
          \right)
         - \eta x \bar x e^{-ix^+} \right)  \nn
   &= {\cal N}^-_{j,\eta} e^{ ijx^+ -  2 i \eta x^- + \frac{\eta y y^*}{2} }
      \sum_{m,n} \frac{|\eta|^m  y^{m-  n} }{m!}(i x)^m (-i \bar x)^{n} 
       e^{-i \frac{ (m + n)x^+}{2}}
       L^{m - n}_{ n} (|\eta| y y^*) \label{wf-}~. 
\end{align}
The normalization is fixed as
\begin{align}
  {\cal N}^{\pm}_{j,\eta} &= \frac{1}{\sqrt{\gamma (|\eta|)}} ~,
 & \gamma (x) & = \frac{\Gamma (x)}{\Gamma (1 - x)} ~.
    \label{N}
\end{align}
For ${\cal H}^0_{j,s}$ we have
\begin{align}
 \Psi^0_{j,s} = \exp \left( ijx^+ + \frac{is}{\sqrt2} 
 \left( y e^{i \psi} 
  + y^* e^{ - i \psi}  \right) \right)
    \sum_{n} e^{i n (\theta + x^+) } 
    \label{wf0}
\end{align}
with $2 \psi = \bar \alpha - \alpha$ and 
$\theta = \alpha + \bar \alpha$.

We can see the relation to the basis labeled by eigenstates of 
$F$ and $J$ using the following expansions as
\begin{align}
 \Psi^{\pm}_{j,\eta} (z,\bar z;x,\bar x)
  &= \sum_{n,\bar n = 0}^{\infty} V^{\pm}_{j,\eta;n,\bar n} (z,\bar z) 
    \frac{\left(x\sqrt{|\eta|} \right)^n}{\sqrt{ n!}}
     \frac{\left(\bar x \sqrt{|\eta|} \right)^{\bar n}}
         {\sqrt{\bar n !}}~,\\
\Psi^0_{j,s} (z,\bar z;\alpha,\bar \alpha)
  &= \sum_{n,\bar n = -\infty}^{\infty} V^0_{j,s;n,\bar n}  (z,\bar z) 
    e^{i n \alpha + i\bar n \bar \alpha} ~,
\end{align}
or inversely
\begin{align}
V^{\pm}_{j,\eta;n,\bar n} (z,\bar z) 
 &= \frac{|\eta|^2}{\pi^2} \int d^2 x d^2 \bar x
    e^{-|\eta| x x^* - |\eta| \bar x \bar x^*}
 \Psi^{\pm}_{j,\eta} (z,\bar z;x,\bar x)
    \frac{\left(x^*\sqrt{|\eta|} \right)^n}{\sqrt{ n!}}
     \frac{\left(\bar x^* \sqrt{|\eta|} \right)^{\bar n}}{\sqrt{\bar n !}}
   ~, \label{momentumpm} \\
V^0_{j,s;n,\bar n}  (z,\bar z) 
 &= \frac{1}{(2 \pi)^2} \int_0^{2\pi} d\alpha \int_0^{2\pi} d \bar \alpha 
  \Psi^0_{j,s} (z,\bar z;\alpha,\bar \alpha)
    e^{- i n \alpha - i\bar n \bar \alpha} ~.
    \label{momentum0}
\end{align}

\subsection{Free field realization and spectral flow}
\label{free}

In order to compute physical quantities, it might be useful to use
free field realization introduced in Refs.~\cite{KK,KKL}.%
\footnote{Other type of free field realization was proposed in 
Ref.~\cite{CFS}, which is not used in this paper.}
Using the free fields
\begin{align}
 X^+ (z) X^- (w) &\sim \ln (z-w) ~,
 & Y(z) Y^*(w) &\sim - \ln (z-w) ~,
 \label{freeOPE}
\end{align}
we can rewrite the currents as
\begin{align}
 J &= \partial X^- ~,
&F &= \partial X^+ ~,
&P^+ &= - i \sqrt{2} e^{-i X^+} \partial Y ~,
&P^- &= - i \sqrt{2} e^{i X^+} \partial Y^* ~.
\end{align}
We can check that the OPEs (\ref{H4OPE}) are reproduced by
using \eqref{freeOPE}.
The primary fields in ${\cal H}^{\pm}_{j,\eta}$ can be expressed by
\begin{equation}
 V^{\pm}_{j,\eta,0} 
 = \exp \left( i j X^+ + i\eta X^-\right) \sigma^{\pm}_{\eta} ~,  
\end{equation}
and the action of $P^{\pm}_0$, where
we introduce the twist fields
\begin{align}
 i \partial Y (z) \sigma^+_\eta (w) &\sim 
 (z-w)^{- \eta} \tau^+_\eta (w) ~,
& i \partial Y^* (z) \sigma^+_\eta (w) &\sim 
 (z-w)^{\eta - 1} {\tau^{+}}'_\eta (w) ~, \nn
 i \partial Y (z) \sigma^-_\eta (w) &\sim 
 (z-w)^{- \eta - 1} {\tau^-}'_\eta (w) ~,
& i \partial Y^* (z) \sigma^-_\eta (w) &\sim 
 (z-w)^{\eta} \tau^-_\eta (w) 
\end{align}
with descendant twist fields 
$ \tau ^{\pm}_\eta , {\tau ^{\pm}} ' _\eta $.
The primary fields in ${\cal H}^{0}_{j,\eta}$ are of the form of
the plane wave as
\begin{align}
 \Psi^0_{j,s}
 = \exp \left( ij X^+ + \frac{is}{\sqrt2} 
 \left( Y e^{i \psi} 
  + Y^* e^{ - i \psi}  \right) \right)
    \sum_{n} e^{i n (\theta + X^+) } ~,
    \label{0js}
\end{align}
therefore the computation involving only the primary fields in  
${\cal H}^{\pm}_{j,\eta}$ is the same as in the flat space-time case.

In the current algebra \eqref{H4CR} there is a symmetry to
redefine the currents as $(w \in \bz)$
\begin{align}
  J_n &\to J_n ~,
 &F_n &\to F_n - i w \delta_{n,0} ~,
 &P^+_n &\to P_{n - w} ~,
 &P^-_n &\to P^-_{n + w }~,
\label{spectral flow}
\end{align}
and this operation is called as spectral flow.\footnote{%
See Refs.~\cite{KP,DK} for more detail. 
Also refer to Ref.~\cite{MO1} for $AdS_3$ case.}
We should include the representation of the currents generated 
by the spectral flow because OPEs do not close without including
the fields with flowed representation.
In the free field realization defined above, we can easily express the
primary fields obtained by the spectral flow as 
($w = 0,\pm 1, \pm 2$ for ${\cal H}^{\pm,w}_{j,\eta}$)
\begin{equation}
 V^{\pm,w}_{j,\eta,0} 
 = \exp \left( i j X^+ + i (\eta + w)X^-\right) \sigma^{\pm}_{\eta} ~,  
\end{equation}
and the action of $P^{\mp}_{- w}$. The corresponding vacuum state 
satisfies
\begin{align}
 J_0 \ket{j,\eta, w} &= i j \ket{j,\eta, w} ~,
&F_0 \ket{j,\eta, w} &=  i (\eta+w)  \ket{j,\eta, w} ~,\nn
 P^{\pm}_n \ket{j,\eta,w} &= 0 ~,~({}^\forall n \geq - w) 
&P_n^{\mp} \ket{j,\eta,w} &= 0 ~.~({}^\forall n > w) 
\end{align}
We should note that we restricted the range of $w$ even though we could
take $w \in \bz$ because there are identities among the Hilbert 
spaces 
\begin{align}
{\cal H}^{+,w}_{j,\eta} \cong {\cal H}^{-,w + 1}_{j,1-\eta} ~.
\end{align} 
The primary fields in ${\cal H}^{0,w}_{j,\eta}$ $(w \in \bz)$ are
expressed as
\begin{align}
 \Psi^0_{j,s}
 = \exp \left( ij X^+ +  i w X^- +\frac{is}{\sqrt2} 
 \left( Y e^{ i \psi} 
  + Y^* e^{- i \psi}  \right) \right)
    \sum_{n} e^{i n (\theta + X^+) } ~,
    \label{wjs}
\end{align}
namely, we only include an integer momentum $p^+ = w$.

\subsection{Two and three point functions}

Two and three point functions of the primary fields are computed
in Refs.~\cite{DK,CFS}. 
They used several methods, such as, free field
realizations, the pp-wave limit of $SU(2) \times U(1)$ WZW model
and the conformal bootstrap from the four point function obtained
by solving Knizhnik-Zamolodchikov equations.
Two point functions are only the normalization, and
meaningful information is in three point functions.

The non-trivial two point functions are the following.
One is the two point function between the primary fields
$\Psi^+_{j,\eta}$ and $\Psi^-_{j,\eta}$ as
\begin{align}
\langle \Psi^+_{j_1,\eta_1} (z_1,\bar z_1;x_1,\bar x_1) 
      \Psi^-_{j_2,\eta_2}  (z_2,\bar z_2;x_2,\bar x_2) \rangle 
 = \frac{ | e^{-\eta_1 x_1 x_2}|^2
   \delta (\eta_1 + \eta_2) \delta ( j_1 + j_2 )}
  {|z_1 - z_2|^{4h_1^+}} 
      \label{2ptpm}
\end{align}
with the notation $|f(x)|^2 = f(x)f(\bar x)$.
Another is the two point function for  $\Psi^0_{j,s}$ as
\begin{align}
&\langle \Psi^0_{j_1,s_1} (z_1,\bar z_1;\alpha_1,\bar \alpha_1) 
      \Psi^0_{j_2,s_2}  (z_2,\bar z_2;\alpha_2,\bar \alpha_2)
  \rangle  \nn&\quad
 = \frac{(2\pi)^2 e^{- i (j_1 + j_2)  \theta_2} \delta (\theta_1 -\theta_2)
  \delta (p^c_1 + p^c_2)
   \delta (p^s_1 + p^s_2)
   \sum_{n=0,1} \delta ( j_1 + j_2 - n)}
    {|z_1 - z_2|^{4h^0_1}} ~,
    \label{2pt0}
\end{align} 
where we define 
\begin{align}
p^c &= s \cos \psi ~,
&p^s &= s \sin \psi ~,
\end{align}
so that the momentum conservation in the $\re y$ and $\im y$ 
directions manifest.
Note that the non-trivial case is only when $j_2 + j_1 = 0$ or 
$j_2 = j_1 = \ts{\frac12}$.

The three point functions involving $\Psi^+_{j,\eta}$ or 
$\Psi^-_{j,\eta}$ are of the form
\begin{align} 
&\left\langle \prod_{i=1}^3 \Psi^{a_i}_{J_i} (z_i,\bar z_i;x_i,\bar x_i) 
 \right\rangle \nonumber \\
 &\qquad = \frac{ C^{a_1 a_2 a_3} (J_1,J_2,J_3) 
          D^{a_1 a_2 a_3} (x_1,\bar x_1,x_2,\bar x_2,x_3,\bar x_3)}
  {|z_1 - z_2|^{2(h_1 + h_2 - h_3)}|z_2 - z_3|^{2(h_2 + h_3 - h_1)}
   |z_3 - z_1|^{2(h_3 + h_1 - h_2)} } ~,
   \label{3ptform}
\end{align}
where $a_i=\pm$ and $J_i=(j_i,\eta_i)$.
The functions $D^{a_1 a_2 a_3}$ can be fixed only from the
symmetry. In other words, they are solutions to 
the differential equations coming from the relation \eqref{diffeq}
as 
\begin{align}
 D^{++-} &= | e^{-x_3 (\eta_1 x_1 + \eta_2 x_2)} 
   (x_2 -x_1)^N|^2 \delta (\eta_1 + \eta_2 + \eta_3) 
   \sum_{n=0}^{\infty} \delta (N-n) ~,\\
 D^{+--} &= | e^{ x_1 (\eta_2 x_2 + \eta_3 x_3)} 
   (x_2 -x_3)^{-N}|^2 \delta (\eta_1 + \eta_2 + \eta_3) 
   \sum_{n=0}^{\infty} \delta (N+n) ~,\label{3ptd}
\end{align}
where $N=-j_1 -j_2 -j_3$.
The coefficients $C^{a_1 a_2 a_3} (J_1,J_2,J_3)$ cannot be determined from
the symmetry and must be computed in other methods.
The results are \cite{DK,CFS}
\begin{align}
C^{++-}(J_1,J_2,J_3) &= \frac{1}{\Gamma (1 + N)}
 \left(\frac{\gamma (-\eta_3)}{\gamma (\eta_1) \gamma (\eta_2)}
 \right)^{\frac12 + N} ~, \label{3ptc1}\\
C^{+--}(J_1,J_2,J_3) &= \frac{1}{\Gamma (1 - N)}
 \left( \frac{\gamma (\eta _1)}{\gamma (-\eta_2) \gamma (-\eta_3)}
 \right)^{\frac12 - N} ~. \label{3ptc2}
\end{align}
The OPEs among the primary fields can be read from the above three
point functions, though we should care about the non-trivial
normalization of the two point functions \eqref{2ptpm}, \eqref{2pt0}.%
\footnote{The normalization \eqref{N} may be read from the 
three point functions \eqref{3ptc1}, \eqref{3ptc2}.
The limit of $\lim_{j,\eta \to 0}\Psi^{\pm}_{j,\eta}$ in 
\eqref{wf+}, \eqref{wf-} is $\lim_{j,\eta \to 0}
{\cal N}^{\pm}_{j,\eta} \cdot {\bf 1}$
where ${\bf 1}$ is the identity.
The three point function with one identity must reduce to the
two point function, which may lead to the normalization \eqref{N}.
Another way to fix the normalization may be using the free field
realization as in Refs.~\cite{CFS} and \cite{BDKZ}.}

For the three point functions only involving $\Psi^0_{j,s}$, 
it is not useful to express in the form of \eqref{3ptform}. 
Since the $\Psi^0_{j,s}$ is of the form of plane wave,
we rather write the three point function as
\begin{align}
&\left \langle \prod_{i=1}^3 \Psi^{0}_{j_i,s_i} 
   (z_i,\bar z_i;x_i,\bar x_i) \right\rangle   \label{3pt0} \\
  & \! =  \frac{(2\pi)^3 e^{i N  \theta_3} \delta (\theta_1 -\theta_3)
   \delta (\theta_2 - \theta_3) 
  \delta (p^c_1 + p^c_2 + p^c_3)
   \delta (p^s_1 + p^s_2 + p^s_3)
   \sum_{n=\pm 1,0} \delta (N- n)}
  {|z_1 - z_2|^{2(h^0_1 + h^0_2 - h^0_3)}|z_2 - z_3|^{2(h^0_2 + h^0_3 - h^0_1)}
   |z_3 - z_1|^{2(h^0_3 + h^0_1 - h^0_2)} } ,   \nonumber
\end{align}
where the momentum conservation is manifest. The normalization is fixed
using the fact that the three point function \eqref{3pt0} should reduce to
the two point function \eqref{2pt0} if one operator is the identity
$V^0_{0,0;0,0} (z,\bar z)$ \eqref{momentum0}.


\section{Geometry of Symmetric D-branes}

D-branes are defined by the hypersurface where the boundary of
open string sweeps. The worldsheet for open string can be given
by a disk or an upper half plane, which can be mapped to each other 
by conformal transformation.
Thus we should assign the boundary condition to the
currents in order to preserve the conformal symmetry. 
Here we only consider the boundary conditions which preserve the
half of the current algebra.

As classified in Ref.~\cite{NWDbrane2} there are essentially two types of 
boundary conditions. One type is
\begin{align}
 J &= (-1)^{\epsilon} \bar J ~,
&F &= (-1)^{\epsilon}\bar F ~,
&P^+ &= (-1)^{\epsilon}\bar P^+ ~,
&P^- &= (-1)^{\epsilon}\bar P^- ~.
\label{bc1}
\end{align}
Since we should perform a conformal mapping to the currents in order
to exchange the open and closed string channel (equivalently to 
exchange the time and space coordinates of worldsheet),
there is a phase factor in front of the currents,
where $\epsilon = 0$ and  $\epsilon = 1$ for the open and 
closed string channel, respectively.
The other type is
\begin{align}
 J &= - (-1)^{\epsilon}\bar J ~,
&F &= - (-1)^{\epsilon}\bar F ~,
&P^+ &= - (-1)^{\epsilon}\bar P^- ~,
&P^- &= - (-1)^{\epsilon}\bar P^+ ~.
\label{bc2}
\end{align}
We do not consider the boundary conditions which reduce
to \eqref{bc1} or \eqref{bc2} by using the inner automorphism.

Let us first examine the boundary condition \eqref{bc2}.
Under the boundary condition, a group element at the boundary 
can be transformed by the adjoint transformation
\begin{align}
 &g^{-1}(x^+,x^-,y) g(x_0^+,x_0^-,y_0)g(x^+,x^-,y)   
  \label{adjoint} \\
& \quad
  = g(x_0^+,x_0^- - \ts{\frac12} |y|^2 \sin x_0^+ 
     +  \im (y_0 y^* 
      e^{\frac{i}{2} x^+} \cos \ts{\frac{x_0^+}{2}}) 
     , e^{\frac{i}{2} x^+} 
       (y_0 e^{\frac{i}{2} x^+} - 2 i y \sin \ts{\frac{x^+_0}{2}}))) ~.
      \nonumber
\end{align}
If the parameters $(x^{\pm},y)$ vary,
then the right hand side of \eqref{adjoint} draws a subspace of
the Nappi-Witten background.
The subspace has the symmetry left after taking the boundary condition,
so it must be the geometry of the D-brane corresponding 
to the boundary condition \eqref{bc2}.
This subspace is classified by the conjugacy class of the group
\cite{NWDbrane2}.

For $x_0^+ = 0 \mod 2\pi$,  the equation \eqref{adjoint} leads to 
\begin{align}
g(0,x_0^- \pm \im ( y_0 y^*  e^{\frac{i}{2} x^+} ), y_0 e^{ix^+}) ~.
\end{align}
Therefore we have 0-dimensional instanton for $y_0=0$ and 
cylindrical 2-dimensional brane for $y_0 \neq 0$.
However we do not consider these branes because their metrics are 
degenerated.
For $x_0^+ \neq 0 \mod 2 \pi$, we can rewrite 
\eqref{adjoint} as
\begin{align}
 g( x_0^+, x_0^-  + 
  \ts{\frac14} \cot \ts{\frac{x_0^+}{2}} (|y_0|^2 -|a|^2),a) ~,
\end{align}
thus the corresponding branes have 2-dimensional Euclidean worldvolume 
located at
\begin{align}
 x^+ &= x^+_0 ~,
&x^- &= {x^-}'_0 + \ts{\frac14} \cot \ts{\frac{x_0^+}{2}} |y|^2 ~.
\label{shapeE2}
\end{align}
If $x_0^+ = \pi \mod 2 \pi$, the worldvolume is flat at 
$x^{\pm} = x^{\pm}_0$, otherwise the worldvolume is hyperbolic.

Another boundary condition \eqref{bc1} implies the invariance of the 
geometry under the twisted adjoint action 
$g \to (r \cdot g_L)^{-1} g g_R$ 
where $r$ generates an outer automorphism 
\begin{align}
r\cdot g(x^+,x^-,y) \to g( - x^+, - x^-, - y^*)
\end{align}
as discussed in Ref.~\cite{NWDbrane2}. Since we have
\begin{align}
 &g^{-1}(- x^+,- x^-,- \bar y) g(x_0^+,x_0^-,y_0)g(x^+,x^-,y)   
  = g(x_0^+ + 2 x^+,    \label{tadjoint} \\
 & \quad x_0^- + 2 x^- 
  + \ts{\frac12} \im (\bar y e^{\frac{i}{2}(x_0^+ + x^+)} 
    (y_0 + \bar y_0  + \bar y e^{\frac{i}{2}(x_0^+ + x^+)}))
     , y_0 + 2 \re ( y e^{-\frac{i}{2}(x_0^+ + x^+)} )) ~,
 \nonumber
\end{align}
the corresponding brane is the (2+1) dimensional hypersurface at 
\begin{align}
\im y = \im y_0 \equiv b
\label{shapeD2}
\end{align}
with a real parameter $b$.
In summary, we have Lorentzian D2-branes at \eqref{shapeD2} for the
boundary condition \eqref{bc1} and Euclidean D2-branes at \eqref{shapeE2} 
for the boundary condition \eqref{bc2}.


\section{One Point Functions and Boundary States}

In quantum level the D-branes are expressed by so-called boundary 
states. In the open string channel, the D-branes are defined by the
hypersurface where the ends of the open strings are attached to.
On the other hand, in the closed string channel, the D-branes are
described by the boundary states which have information how 
closed strings couple to the D-branes.
A general boundary state is given by a linear combination of the 
Ishibashi states, 
and the coefficients 
can be read from the disk (or upper half plane)
one point functions of primary fields.
In this section, we compute the one point functions with boundary 
conditions \eqref{bc1}, \eqref{bc2} and construct the boundary states
for the Lorentzian D2-branes and Euclidean D2-brane.

\subsection{Lorentzian D2-branes}

Let us first consider the Lorentzian D2-branes corresponding to the 
boundary condition \eqref{bc1}. The boundary condition leads to
\begin{align}
 ({\cal D}_x^{J,F,\pm} + \bar{\cal D}_{\bar x}^{J,F,\pm}) 
\langle \Psi^a_{J} (z,\bar z ; \alpha, \bar \alpha) \rangle 
= 0 ~,
\label{diffopen}
\end{align}
because of the relation \eqref{diffeq}.
The condition ${\cal D}_x^F+\bar {\cal D}_{\bar x}^F = 0$ means that
only the one point functions of the states in ${\cal H}^0_{j,s}$ have 
no-trivial value (even including the spectral flowed states). 
This is reasonable since D-brane parallel to the light-cone direction 
couples to the closed strings with vanishing light-cone momentum.
The other conditions restrict the form of the one point function to
\begin{align}
 \langle \Psi^0_{j,s} (z,\bar z ; \alpha, \bar \alpha) \rangle 
   &= \frac{U_{j,s} (\alpha,\bar \alpha)}{|z-\bar z|^{2 h^0}} ~,
   & U_{j,s}(\alpha,\bar \alpha) = f_{j,s}
   \delta (p^c) e^{- i j \theta} \sum_{n=0,1} \delta (2j -n)~.
   \label{1ptD2}
\end{align}
This form may be also derived from the free field realization 
\eqref{0js}. In fact, it is the one point function with the
Newmann and Dirichlet boundary conditions (in the open string terms)
for the the $\re y$ and $\im y$ directions, respectively.

In the following, it is convenient to define
\begin{align}
 \hat \Psi^0_{\hat j,s} (z,\bar z ; \psi) 
 \equiv \frac{1}{2\pi} \int_0^{2\pi} d\theta e^{i \hat j \theta}
 \Psi^0_{\hat j,s} (z,\bar z ; \alpha, \bar \alpha) ~,
 \label{newnorm}
\end{align}
with $\hat j = 0,1/2$, and compute its one point function
\begin{align}
 \langle \hat \Psi^0_{\hat j,s} (z,\bar z ; \psi) \rangle 
   &= \frac{ f_{\hat j,s} 
 \delta (p^c)}{|z-\bar z|^{2 h^0}} ~.
 \label{hat1ptD2}
\end{align}
We also fix the normalization of the one point function such as
\begin{align}
\langle V^0_{0,0;0,0} (z,\bar z) \rangle = 1 ~,
\label{1ptnorm}
\end{align}
namely, the one point function of the identity is set to be one.

In order to compute the coefficient $f_{\hat j,s}$, we utilize the
conformal bootstrap method \cite{CL,Lewellen,PSS}.
We consider the following two point function on the upper half plain
\begin{align}
 \langle \hat \Psi^0_{\hat j_1,s_1} (z_1,\bar z_1 ; \psi_1) 
  \hat \Psi^0_{\hat j_2,s_2} (z_2,\bar z_2 ; \psi_2) \rangle 
 ~.
 \label{2ptD2}
\end{align}
This function can be computed in two ways. 
One way is using the OPE of the inserted closed strings
\begin{align}
 & \hat \Psi^0_{\hat j_1,s_1} (z,\bar z ; \psi_1) 
  \hat \Psi^0_{\hat j_2,s_2} (w, \bar w ; \psi_2) 
 \sim  \sum_{\hat j_3 - \hat j_1 - \hat j_2 = 0, \pm 1}\int d p^c_3 d p^s_3
 \frac{\hat C^{000}_{123} 
   \hat \Psi^0_{\hat j_3,s_3} (w,\bar w ; \psi_3)}
   {|z - w|^{2(h_1 + h_2 - h_3)}} 
\end{align}
up to the contribution of descendants, where
the three point coefficient is given by
\begin{align}
 \hat C^{000}_{123} =  
  \delta (p^c_1 + p^c_2 - p^c_3) 
   \delta (p^s_1 + p^s_2 - p^s_3) ~.
\end{align}
Another way is
to use the OPEs between the bulk operators and boundary operators.
If the boundary operator is the identity, 
then the coefficient
of the OPE is the same as the one point function \eqref{hat1ptD2}
under the normalization \eqref{1ptnorm}.
Comparing the two ways of computing the two point function 
\eqref{2ptD2}, we have the equality
\begin{align}
 \int d p^c_3 d p^s_3
   \hat C^{000}_{123} f_{\hat j_3,s_3} \delta (p^c_3) 
  {\cal F}^{12}_{\bar 2 \bar 1} (z) =
    f_{\hat j_1,s_1}  \delta (p^c_1)
    f_{\hat j_2,s_2}  \delta (p^c_2) 
  {\cal F}^{1 \bar 1}_{2 \bar 2} (1-z)
   \label{cbootstrap}
\end{align}
as long as the propagating boundary operator is the identity.%
\footnote{We changed the normalization \eqref{newnorm}
such that the identity appears as the propagating boundary operator with
the correct normalization.}
We denote ${\cal F} (z)$ as the four point conformal blocks 
\begin{align}
 {\cal F}^{12}_{\bar 2 \bar 1} (z) =  {\cal F}^{1 \bar 1}_{2 \bar 2} (1-z)
  = z^{p^c_1 p^c_2 + p^s_1 p^s_2} 
   (1-z)^{- (p^c_1)^2 + (p^s_1)^2 } ~,
\end{align}
which correspond to the choice of 
$z_1 = z,z_2=0,\bar z_2=\infty, \bar z_1=1$. 
Solution to the equation \eqref{cbootstrap} is\footnote{
Note that $\sin \psi$ only set the sign in front of $s$
because of the delta function $\delta (s \cos \psi)$ in 
\eqref{hat1ptD2}.}
\begin{align}
 f_{\hat j,s} = \exp (i p^s b) 
 \label{2brane}
\end{align}
with a parameter $b$.
We should notice that the normalization is consistent with 
\eqref{1ptnorm}.
In summary we have the non-trivial one-point function of the
closed strings for ${\cal H}^0_{j,s}$ as
\begin{align}
 \langle \Psi^0_{j,s} (z,\bar z ; \alpha, \bar \alpha) \rangle _b
   &= \frac{e^{i p^s b - i j \theta}  \delta (p^c) 
    \sum_{n=0,1} \delta (2j -n)}{|z-\bar z|^{2 h^0}}
   \label{1ptD2sum}
\end{align}
for the Lorentzian D2-branes corresponding to 
the boundary condition \eqref{bc1}.

The boundary state for the Lorentzian D2-brane is defined to
reproduce the one point function \eqref{1ptD2sum}, namely
\begin{align}
 \langle \langle b | j,s,\alpha, \bar \alpha \rangle 
 = | z - \bar z |^{2 h^0}   
 \langle \Psi^0_{j,s} (z,\bar z ; \alpha, \bar \alpha) \rangle _b
 ~, \label{bsD2}
\end{align}
where $\ket{j,s,\alpha,\bar \alpha}$ is closed string state corresponding
to the primary fields $\Psi^0_{j,s} (\alpha, \bar \alpha)$.
Noticing that classically 
$\langle j,s,\alpha,\bar \alpha | x^+ , x^- , y \rangle = \Psi^0_{j,s}$
in \eqref{wf0}, we have 
\begin{align} 
 \langle \langle b | x^+ , x^- , y \rangle &\sim 
\frac{1}{(2\pi)^2} \int_{-\frac12}^{\frac12} d j \int_0^{\infty} sds
 \int_0^{2\pi} d \alpha \int _0^{2\pi} d \bar \alpha  
 \langle \Psi^0_{-j,s} (\bar \alpha, \alpha ) \rangle  
 \Psi^0_{j,s} ( \alpha , \bar \alpha ) \nn &\sim
 \delta (\im y - b) ~.
\end{align}
Therefore, we can conclude that the boundary state \eqref{bsD2} 
reproduces the classical geometry of the D2-branes \eqref{shapeD2}, 
where we identify the parameter $b$ in \eqref{2brane} as the 
position of the D2-brane.

\subsection{Euclidean D2-branes}

The boundary condition \eqref{bc2} also leads to the condition
\begin{align}
 ({\cal D}_x^{J,F,\pm} - \bar{\cal D}_{\bar x}^{J,F,\mp}) 
\langle \Psi^a_{J} (z,\bar z ; \alpha, \bar \alpha) \rangle 
= 0 ~,
\label{diffopen2}
\end{align}
which is similar to \eqref{diffopen}.
In this case the condition ${\cal D}_x^F - \bar{\cal D}_{\bar x}^F = 0$ 
gives no constraint on $\eta$,  which implies that all the closed 
strings couple to the brane contrary to the previous case.
The other conditions restrict the form of one point functions as
before.

We first consider the primary fields in ${\cal H}^0_{j,s}$,
whose one point function is of the form%
\footnote{The conditions \eqref{diffopen2} allow the form of
$f_{j,s} \delta (\theta)$ other than \eqref{1pte0}, 
however we do not use it because it is
not consistent with the classical picture \eqref{shapeE2}.}
\begin{align}
 \langle \Psi^0_{j,s} (z,\bar z ; \alpha, \bar \alpha) \rangle 
   &= f_j (\theta) \frac{\delta (s)}{s}  ~.
   \label{1pte0}
\end{align}
Just like the previous case it is useful to define
\begin{align}
 \hat \Psi^0_{j,n} (z,\bar z) = \frac{1}{(2\pi)^2}
   \int_0^{2\pi} d \psi \int_0^{2\pi} d \theta e^{-i n \theta} 
   \Psi^0_{j,0} (z,\bar z ; \alpha, \bar \alpha) ~,
   \label{newop2}
\end{align}
and compute its one point function
\begin{align}
 \langle \hat \Psi^0_{j,n} (z,\bar z) \rangle 
   &=  \hat f_{j,n}  ~.
\end{align}
We set the normalization as \eqref{1ptnorm} 
(even though the normalization of the one point function
is different from the previous one in general).
Because the OPE involving \eqref{newop2} is given by 
\begin{align}
 \hat \Psi^0_{j_1,n_1} (z,\bar z) \sim
  \hat \Psi^0_{j_2,n_2} (w, \bar w) \hat 
  \Psi^0_{j_1 + j_2 + m,n_1 + n_2 - m} (w,\bar w)
    + \cdots ~
\end{align}
($m$ is chosen so that $-1/2 < j_1 + j_2 + m \leq 1/2$),
we have the conformal bootstrap constraint 
\begin{align} 
  \hat f_{j_1 + j_2 + m,n_1 + n_2 - m}  
   = \hat f_{j_1,n_1} \hat f_{j_2,n_2}
\end{align}
as in \eqref{cbootstrap}.
Solution to the constraint is given by
\begin{align}
 \hat f_{j,n} = \exp ( i(j + n) x^+_0 ) ~.
 \label{hatf}
\end{align}
Since the wave function corresponding to \eqref{newop2} is
\begin{align}
 \hat \Psi^0_{j,n} = \exp ( i (j + n) x^+ ) ~,
\end{align}
the one point function implies that the brane is at $x^+ = x^+_0$,
which is consistent with the classical analysis \eqref{shapeE2}.

Let us move to the closed strings in ${\cal H}^{\pm}_{j,\eta}$. 
{}From the conditions \eqref{diffopen2}, the form of the one point
function is restricted as
\begin{align}
 \langle \Psi^{\pm}_{j,\eta} (z,\bar z ; x, \bar x) \rangle 
   &= \frac{U^{\pm}_{j,\eta} e^{|\eta |x \bar x}}
           {|z - \bar z|^{2h^{\pm}}} ~.
   \label{1ptepm}
\end{align}
As before we compute the two point function
\begin{align}
 \langle \Psi^+_{j_1,\eta_1} (z_1,\bar z_1 ; x_1 , \bar x_1) 
  \Psi^-_{ j_2,\eta_2} (z_2,\bar z_2 ; x_2 , \bar x_2) \rangle 
 \label{2ptepm}
\end{align}
in order to obtain constraints for $U^{\pm}_{j,\eta}$.
Here we assume $\eta_1 + \eta_2 > 0$.%
\footnote{The other cases $\eta_1 + \eta_2 < 0$ and 
$\eta_1 + \eta_2 = 0$ give other constraints. 
The former case leads to the similar constraint, 
and the latter case gives the relation 
between \eqref{1pte0} and \eqref{1ptepm}.
It is very important to check also in these cases 
although we will not do it in this paper. Another important constraint
may come from the two point function with $\Psi^0_{j,s}$ and 
$\Psi^{\pm}_{j,\eta}$.}
Computing two ways and comparing the both, we obtain
\begin{align}
 \sum_{n = 0}^{\infty} C^{+--} 
(j_1,\eta_1;j_2,\eta_2;-j_1-j_2+n,-\eta_1-\eta_2)
 U^+_{j_1 + j_2 - n,\eta_1 + \eta_2} 
 {\cal F}^{12}_{\bar 2 \bar 1}(n,z,x)\nn
  = \int_0^{\infty} s ds C_B^{+0}(j_1,\eta_1;s,0)
   C_B^{-0}(j_2,\eta_2;s,0)
    {\cal F}^{1 \bar 1}_{2 \bar 2} (s,1-z,x) ~,
    \label{const1}
\end{align}
where $C_B^{\pm 0}$ represents the bulk-boundary two point function with
$C_B^{\pm 0}(j,\eta;0,0) = U^{\pm}_{j,\eta}$.
We used
the conformal blocks ${\cal F}^{12}_{\bar 2 \bar 1}(n,z,x)$
and ${\cal F}^{1 \bar 1}_{2 \bar 2} (s,1-z,x)$ defined
in Ref.~\cite{DK} for $\langle +-+- \rangle$ amplitudes, 
whose intermediate states belong to ${\cal H}^+_{j,\eta}$ 
and ${\cal H}^0_{j,s}$, respectively.\footnote{
The definition of the identity state in Ref. \cite{DK} 
is not the same as the one in this paper,
and hence the normalization of the one point
functions \eqref{1pte0} and \eqref{1ptepm} may be different.}
The transformation of the conformal blocks can be computed as
$(  N = -j_1 - j_2 - j_3 - j_4 )$
\begin{align}
 {\cal F}^{12}_{3 4}(n,z,x) = 
 \int_0^{\infty} s ds (C_n^N (s))^{-1} 
  {\cal F}^{1 4}_{3 2} (s,1-z,x ) ~,
  \label{transfer}
\end{align}
where
\begin{align}
(C_n^N (s))^{-1} = \frac{(-1)^{N/2} n! \Xi^{N + 1}}
                        {A^{n + N + 1} B^n} 
 \left( \frac{s^2}{2} \right)^{\frac{N}{2}}  
  e^{\frac{s^2}{2} ( \psi (|\eta_2| ) + \psi ( 1 - \eta_1 ) - 2 \psi (1) )}
  L_n^N \left(\frac{s^2}{2} \Xi \right)  ~.
\end{align}
Here we have used $\psi(x)=\frac{d}{dx} \ln \Gamma (x)$ and defined
\begin{align}
 \Xi &= \frac{\pi \sin \pi (\eta_1 + \eta_2 )}
             {\sin \pi \eta_1 \sin \pi |\eta_2|} ~,
  &A &=  \frac {\Gamma (1 + \eta_2) \Gamma (\eta_1)}
               {\Gamma (\eta_1 + \eta_2) }  ~,
  &B &=  \frac{\Gamma ( 1 - \eta_1 - \eta_2 )}
             {\Gamma (1 - \eta_1 ) \Gamma ( - \eta_2 )} ~.
\end{align}
With the help of the transformation matrix \eqref{transfer} with
$N=0$ and $s=0$, the relation \eqref{const1} leads to
\begin{align} 
 \sum_{n=0}^{\infty} \sqrt{\Xi} U^+_{j_1 + j_2 - n,\eta_1 + \eta_2}
  = U^+_{j_1,\eta_1} U^-_{j_2,\eta_2} ~.
\end{align}
Solutions to the above equation are given by
\begin{align}
 U^{\pm}_{j,\eta} = \sqrt{\frac{\pi}{\sin \pi |\eta|} }
   \sum_{n=0}^{\infty} e^{i ( j \pm n ) x^+_0 -2 i \eta x^-_0 } 
    =  \sqrt{\frac{\pi}{\sin \pi |\eta|} } 
       \frac{1}{1 - e^{\pm i x_0^+}} e^{i j x^+_0 -2 i \eta x^-_0}
\label{Upm}
\end{align}
with parameters $x^-_0$ and $x^+_0 \neq 0 \mod 2 \pi$.

We can construct the boundary state like \eqref{bsD2} and
compute the coupling with localized closed strings as
\begin{align}
 \langle \langle x^+_0, x^-_0 | x^+ , x^- , y \rangle &\sim
 \int_{-\infty}^{\infty} d j \int_{-1}^1 d \eta 
  \frac{\eta ^2}{\pi ^2} \int d^2 x  d^2 \bar x 
    \langle \Psi^{\mp}_{ - j, - \eta} ( - x^* , - \bar x^* ) \rangle 
            \Psi^{\pm}_{j,\eta} (x,\bar x)\nn
  &\sim \int_{-1}^1 d \eta 
   \frac{\pi \delta (x^+ - x^+_0) \Gamma (1-|\eta|)}{2 \sin ^2 \ts{\frac{x_0^+}{2}} }
    e^{- 2 i \eta (x^- - x^-_0 + \ts{\frac14} \cot \ts{\frac{x^+_0}{2}} |y|^2 ) }~.
\end{align}
To see its classical behavior, it is convenient to reintroduce
$\alpha'=2$ and rewrite such as 
$\sqrt{\frac{\alpha'}{2}} \eta, \sqrt{\frac{2}{\alpha'}}x^-$. 
Then, in the classical limit $\alpha' \to \infty$, the small $\eta$ 
region is expanded, and the integration gives delta function 
$\delta (x^- - x^-_0 + \ts{\frac14} \cot \ts{\frac{x^+_0}{2}} |y|^2)$.
Identifying the two free parameters $x^{\pm}_0$ as those of 
\eqref{shapeE2}, we successfully reproduce the classical geometry 
of the Euclidean D2-brane from the boundary state.

For the couplings with the flowed states, it is useful to use the
free field realization as mentioned in section \ref{free}.
Then, we can see from the above results that
$X^-$ field satisfies the Dirichlet boundary 
condition (in the open string terms) 
and the one point function is proportional to
$\exp (-2 i x^-_0 (\eta + w) )$.
Therefore, the one point functions are obtained as
\begin{align}
 \langle \Psi^{0,w}_{j,s} (z,\bar z ; \alpha, \bar \alpha) \rangle 
   &= \sum_{n} e^{ i n \theta + i (j + n) x^+_0 -2 i w x^-_0 } 
      \frac{\delta (s)}{s} 
   \label{1pte0sum}
\end{align}
for ${\cal H}^{0,w}_{j,s}$  with $w \in \bz$ and
\begin{align}
 \langle \Psi^{\pm,w}_{j,\eta} (z,\bar z ; x, \bar x) \rangle 
   &= \sqrt{\frac{\pi}{\sin \pi |\eta|} }
   \sum_{n=0}^{\infty} e^{ i ( j \pm n ) x^+_0 -2 i (\eta + w) x^-_0}
     \frac{ e^{|\eta |x \bar x}}{|z - \bar z|^{2 h^{\pm}}}
   \label{1ptepmsum}
\end{align}
for ${\cal H}^{\pm,w}_{j,\eta}$ with $w = 0,\pm 1,\pm 2,\cdots$.

\section{Conclusion}
\label{conclusion}

We investigated D-branes in the Nappi-Witten model \cite{NW}, 
which is a WZW model associated with 4 dimensional Heisenberg 
group $H_4$.
Only the symmetric D-branes have been considered, whose
boundary conditions preserve the half of the current algebra 
\eqref{bc1}, \eqref{bc2}.
These branes are classified by the (twisted) conjugacy classes
\cite{NWDbrane2},
and we have Lorentzian D2-brane at \eqref{shapeD2} 
and Euclidean D2 brane at \eqref{shapeE2}. 
In this paper we computed the one point functions of closed
strings on the upper half plane. 
It is rather difficult to calculate the one point function directly,
so we computed them by utilizing conformal bootstrap 
constraints \cite{CL,Lewellen,PSS}.
Two point function on the upper half plane can be mapped to 
four point function on the full plane, whose conformal block 
is obtained by solving Knizhnik-Zamolodchikov equations \cite{DK}.
Computing the two point functions in two ways and comparing the
both, we have constraints to the one point functions.
Solving the constraints, we obtained the one point functions 
\eqref{1ptD2sum} for the Lorentzian D2-branes and 
\eqref{1pte0sum}, \eqref{1ptepmsum} for the Euclidean D2-branes.
We constructed the boundary states based on the one point functions 
and checked that the classical limits reproduce the geometry
of the corresponding D-branes.

The methods we used to obtain the
boundary states in the Nappi-Witten model might be useful to
investigate some time-dependent D-branes such as D-branes with
rolling tachyon \cite{Sen1,Sen2}.
This is because the Nappi-Witten model is solvable and its
target space-time is Lorentzian, and hence we do not need to perform 
analytic continuation of Euclidean results, 
which may give rise to difficulty.
An interesting case may be the D-branes in $AdS_3$,
where the boundary states in the Euclidean $AdS_3$ 
is given in Refs.~\cite{AdSDbrane1,AdSDbrane2}.%
\footnote{D-branes in Lorentzian $SL(2,\br)/U(1)$ 
WZW model (which is a coset of the Lorentzian $AdS_3$ model) 
are also worth to study because of the non-trivial
time-dependence of D-branes \cite{Y}.
In order to construct the boundary states, we have to perform
analytic continuation to their Euclidean counterparts, which
are analyzed in Ref. \cite{RS}.}
Since the pp-wave limit of $AdS_3 \times S^3$ is very similar
to our model, our results may give insights in the construction of 
boundary states in the Lorentzian $AdS_3$
(see Ref.~\cite{NWDbrane3} for a classical argument).
The most difficult task with Lorentzian signature is to perform 
the modular transformation because the worldsheet should be also 
Lorentzian. Even in our case, it is difficult to read the
open string spectrum\footnote{It might be possible to directly
analyze the open strings in the Nappi-Witten model. In particular,
we may obtain two and three point functions by utilizing conformal
bootstrap as in Ref.~\cite{AdSDbrane2}.
} by using the modular transformation of one-loop
amplitude, even though the cylinder amplitude in the closed string 
channel is easy to compute using our boundary states.\footnote{We
may be able to perform the modular transformation only in the 
light-cone gauge as in Ref.~\cite{Green1}.}
It would be interesting to try to manage the difficulty.

\section*{Acknowledgement}

I would like to thank S.-J.~Rey, Y.~Satoh, Y.~Sugawara, T.~Takayanagi, 
H.~Takayanagi and A.~Yamaguchi for useful discussions.

\baselineskip=13pt
\providecommand{\href}[2]{#2}\begingroup\raggedright\endgroup


\begin{thebibliography}{10}

\bibitem{BMN}
D.~Berenstein, J.~M. Maldacena, and H.~Nastase, ``Strings in flat space and pp
  waves from {${\cal N} = 4$} super {Yang} {Mills},'' {\em JHEP} {\bf 04}
  (2002) 013,
\href{http://www.arXiv.org/abs/hep-th/0202021}{{\tt hep-th/0202021}}.

\bibitem{mspp-wave1}
M.~Blau, J.~Figueroa-O'Farrill, C.~Hull, and G.~Papadopoulos, ``A new maximally
  supersymmetric background of {IIB} superstring theory,'' {\em JHEP} {\bf 01}
  (2002) 047,
\href{http://www.arXiv.org/abs/hep-th/0110242}{{\tt hep-th/0110242}}.

\bibitem{pp-wave1}
R.~R. Metsaev, ``Type {IIB} {Green-Schwarz} superstring in plane wave
  {Ramond-Ramond} background,'' {\em Nucl. Phys.} {\bf B625} (2002) 70--96,
\href{http://www.arXiv.org/abs/hep-th/0112044}{{\tt hep-th/0112044}}.

\bibitem{pp-wave2}
R.~R. Metsaev and A.~A. Tseytlin, ``Exactly solvable model of superstring in
  plane wave {Ramond-Ramond} background,'' {\em Phys. Rev.} {\bf D65} (2002)
  126004,
\href{http://www.arXiv.org/abs/hep-th/0202109}{{\tt hep-th/0202109}}.

\bibitem{ppdbrane2}
A.~Dabholkar and S.~Parvizi, ``{D$p$} branes in pp-wave background,'' {\em
  Nucl. Phys.} {\bf B641} (2002) 223--234,
\href{http://www.arXiv.org/abs/hep-th/0203231}{{\tt hep-th/0203231}}.

\bibitem{KNS}
A.~Kumar, R.~R. Nayak, and S.~Siwach, ``D-brane solutions in pp-wave
  background,'' {\em Phys. Lett.} {\bf B541} (2002) 183--188,
\href{http://www.arXiv.org/abs/hep-th/0204025}{{\tt hep-th/0204025}}.

\bibitem{ST1}
K.~Skenderis and M.~Taylor, ``Branes in {AdS} and pp-wave spacetimes,'' {\em
  JHEP} {\bf 06} (2002) 025,
\href{http://www.arXiv.org/abs/hep-th/0204054}{{\tt hep-th/0204054}}.

\bibitem{BMZ}
P.~Bain, P.~Meessen, and M.~Zamaklar, ``Supergravity solutions for {D-branes}
  in {Hpp-wave} backgrounds,'' {\em Class. Quant. Grav.} {\bf 20} (2003)
  913--934,
\href{http://www.arXiv.org/abs/hep-th/0205106}{{\tt hep-th/0205106}}.

\bibitem{Hikida}
Y.~Hikida and S.~Yamaguchi, ``D-branes in pp-waves and massive theories on
  worldsheet with boundary,'' {\em JHEP} {\bf 01} (2003) 072,
\href{http://www.arXiv.org/abs/hep-th/0210262}{{\tt hep-th/0210262}}.

\bibitem{ST2}
K.~Skenderis and M.~Taylor, ``Open strings in the plane wave background. {I}:
  Quantization and symmetries,'' {\em Nucl. Phys.} {\bf B665} (2003) 3--48,
\href{http://www.arXiv.org/abs/hep-th/0211011}{{\tt hep-th/0211011}}.

\bibitem{ST3}
K.~Skenderis and M.~Taylor, ``Open strings in the plane wave background. {II}:
  Superalgebras and spectra,'' {\em JHEP} {\bf 07} (2003) 006,
\href{http://www.arXiv.org/abs/hep-th/0212184}{{\tt hep-th/0212184}}.

\bibitem{ppdbrane1}
M.~Billo and I.~Pesando, ``Boundary states for {GS} superstrings in an {Hpp}
  wave background,'' {\em Phys. Lett.} {\bf B536} (2002) 121--128,
\href{http://www.arXiv.org/abs/hep-th/0203028}{{\tt hep-th/0203028}}.

\bibitem{Green1}
O.~Bergman, M.~R. Gaberdiel, and M.~B. Green, ``D-brane interactions in type
  {IIB} plane-wave background,'' {\em JHEP} {\bf 03} (2003) 002,
\href{http://www.arXiv.org/abs/hep-th/0205183}{{\tt hep-th/0205183}}.

\bibitem{Green2}
M.~R. Gaberdiel and M.~B. Green, ``The {D-instanton} and other supersymmetric
  {D-branes} in {IIB} plane-wave string theory,'' {\em Ann. Phys.} {\bf 307}
  (2003) 147--194,
\href{http://www.arXiv.org/abs/hep-th/0211122}{{\tt hep-th/0211122}}.

\bibitem{Green3}
M.~R. Gaberdiel, M.~B. Green, S.~Schafer-Nameki, and A.~Sinha, ``Oblique and
  curved {D-branes} in {IIB} plane-wave string theory,'' {\em JHEP} {\bf 10}
  (2003) 052,
\href{http://www.arXiv.org/abs/hep-th/0306056}{{\tt hep-th/0306056}}.

\bibitem{NW}
C.~R. Nappi and E.~Witten, ``A {WZW} model based on a nonsemisimple group,''
  {\em Phys. Rev. Lett.} {\bf 71} (1993) 3751--3753,
\href{http://www.arXiv.org/abs/hep-th/9310112}{{\tt hep-th/9310112}}.

\bibitem{KP}
E.~Kiritsis and B.~Pioline, ``Strings in homogeneous gravitational waves and
  null holography,'' {\em JHEP} {\bf 08} (2002) 048,
\href{http://www.arXiv.org/abs/hep-th/0204004}{{\tt hep-th/0204004}}.

\bibitem{HS1}
Y.~Hikida and Y.~Sugawara, ``Superstrings on pp-wave backgrounds and symmetric
  orbifolds,'' {\em JHEP} {\bf 06} (2002) 037,
\href{http://www.arXiv.org/abs/hep-th/0205200}{{\tt hep-th/0205200}}.

\bibitem{LM}
O.~Lunin and S.~D. Mathur, ``Rotating deformations of {$AdS_3 \times S_3$}, the
  orbifold {CFT} and strings in the pp-wave limit,'' {\em Nucl. Phys.} {\bf
  B642} (2002) 91--113,
\href{http://www.arXiv.org/abs/hep-th/0206107}{{\tt hep-th/0206107}}.

\bibitem{GMS}
J.~Gomis, L.~Motl, and A.~Strominger, ``pp-wave/{CFT$_2$} duality,'' {\em JHEP}
  {\bf 11} (2002) 016,
\href{http://www.arXiv.org/abs/hep-th/0206166}{{\tt hep-th/0206166}}.

\bibitem{HS2}
Y.~Hikida and Y.~Sugawara, ``Superstring vacua of 4-dimensional pp-waves with
  enhanced supersymmetry,'' {\em JHEP} {\bf 10} (2002) 067,
\href{http://www.arXiv.org/abs/hep-th/0207124}{{\tt hep-th/0207124}}.

\bibitem{Narain}
E.~Gava and K.~S. Narain, ``Proving the pp-wave/{CFT$_2$} duality,'' {\em JHEP}
  {\bf 12} (2002) 023,
\href{http://www.arXiv.org/abs/hep-th/0208081}{{\tt hep-th/0208081}}.

\bibitem{Hikidaphd}
Y.~Hikida, ``Superstrings on {NSNS} pp-waves and their {CFT} duals,''
\href{http://www.arXiv.org/abs/hep-th/0303222}{{\tt hep-th/0303222}}.

\bibitem{DK}
G.~D'Appollonio and E.~Kiritsis, ``String interactions in gravitational wave
  backgrounds,'' {\em Nucl. Phys.} {\bf B674} (2003) 80--170,
\href{http://www.arXiv.org/abs/hep-th/0305081}{{\tt hep-th/0305081}}.

\bibitem{CFS}
Y.-K.~E. Cheung, L.~Freidel, and K.~Savvidy, ``Strings in gravimagnetic
  fields,'' {\em JHEP} {\bf 02} (2004) 054,
\href{http://www.arXiv.org/abs/hep-th/0309005}{{\tt hep-th/0309005}}.

\bibitem{BDKZ}
M.~Bianchi, G.~D'Appollonio, E.~Kiritsis, and O.~Zapata, ``String amplitudes in
  the {Hpp-wave} limit of {$AdS_3 \times S^3$},'' {\em JHEP} {\bf 04} (2004)
  074,
\href{http://www.arXiv.org/abs/hep-th/0402004}{{\tt hep-th/0402004}}.

\bibitem{NWDbrane1}
S.~Stanciu and A.~A. Tseytlin, ``D-branes in curved spacetime: {Nappi-Witten}
  background,'' {\em JHEP} {\bf 06} (1998) 010,
\href{http://www.arXiv.org/abs/hep-th/9805006}{{\tt hep-th/9805006}}.

\bibitem{NWDbrane2}
J.~M. Figueroa-O'Farrill and S.~Stanciu, ``More {D-branes} in the
  {Nappi-Witten} background,'' {\em JHEP} {\bf 01} (2000) 024,
\href{http://www.arXiv.org/abs/hep-th/9909164}{{\tt hep-th/9909164}}.

\bibitem{NWDbrane3}
S.~Stanciu and J.~Figueroa-O'Farrill, ``Penrose limits of {Lie} branes and a
  {Nappi-Witten} braneworld,'' {\em JHEP} {\bf 06} (2003) 025,
\href{http://www.arXiv.org/abs/hep-th/0303212}{{\tt hep-th/0303212}}.

\bibitem{Takayanagi}
H.~Takayanagi and T.~Takayanagi, ``Open strings in exactly solvable model of
  curved space-time and pp-wave limit,'' {\em JHEP} {\bf 05} (2002) 012,
\href{http://www.arXiv.org/abs/hep-th/0204234}{{\tt hep-th/0204234}}.

\bibitem{Michishita}
Y.~Michishita, ``D-branes in {NSNS} and {RR} pp-wave backgrounds and
  {S}-duality,'' {\em JHEP} {\bf 10} (2002) 048,
\href{http://www.arXiv.org/abs/hep-th/0206131}{{\tt hep-th/0206131}}.

\bibitem{Panigrahi1}
A.~Biswas, A.~Kumar, and K.~L. Panigrahi, ``{$p - p'$} branes in pp-wave
  background,'' {\em Phys. Rev.} {\bf D66} (2002) 126002,
\href{http://www.arXiv.org/abs/hep-th/0208042}{{\tt hep-th/0208042}}.

\bibitem{Panigrahi2}
S.~F. Hassan, R.~R. Nayak, and K.~L. Panigrahi, ``D-branes in the {NS5}
  near-horizon pp-wave background,''
\href{http://www.arXiv.org/abs/hep-th/0312224}{{\tt hep-th/0312224}}.

\bibitem{Hpp}
G.~D'Appollonio and E.~Kiritsis, ``D-branes and {BCFT} in {Hpp-wave}
  backgrounds,''
\href{http://www.arXiv.org/abs/hep-th/0410269}{{\tt hep-th/0410269}}.

\bibitem{KK}
E.~Kiritsis and C.~Kounnas, ``String propagation in gravitational wave
  backgrounds,'' {\em Phys. Lett.} {\bf B320} (1994) 264--272,
\href{http://www.arXiv.org/abs/hep-th/9310202}{{\tt hep-th/9310202}}.

\bibitem{KKL}
E.~Kiritsis, C.~Kounnas, and D.~Lust, ``Superstring gravitational wave
  backgrounds with space-time supersymmetry,'' {\em Phys. Lett.} {\bf B331}
  (1994) 321--329,
\href{http://www.arXiv.org/abs/hep-th/9404114}{{\tt hep-th/9404114}}.

\bibitem{MO1}
J.~M. Maldacena and H.~Ooguri, ``Strings in {$AdS_3$} and the {$SL(2,\br)$}
  {WZW} model. {I}: The spectrum,'' {\em J. Math. Phys.} {\bf 42} (2001)
  2929--2960,
\href{http://www.arXiv.org/abs/hep-th/0001053}{{\tt hep-th/0001053}}.

\bibitem{CL}
J.~L. Cardy and D.~C. Lewellen, ``Bulk and boundary operators in conformal
  field theory,'' {\em Phys. Lett.} {\bf B259} (1991)
274--278.

\bibitem{Lewellen}
D.~C. Lewellen, ``Sewing constraints for conformal field theories on surfaces
  with boundaries,'' {\em Nucl. Phys.} {\bf B372} (1992)
654--682.

\bibitem{PSS}
G.~Pradisi, A.~Sagnotti, and Y.~S. Stanev, ``Completeness conditions for
  boundary operators in 2d conformal field theory,'' {\em Phys. Lett.} {\bf
  B381} (1996) 97--104,
\href{http://www.arXiv.org/abs/hep-th/9603097}{{\tt hep-th/9603097}}.

\bibitem{Sen1}
A.~Sen, ``Rolling tachyon,'' {\em JHEP} {\bf 04} (2002) 048,
\href{http://www.arXiv.org/abs/hep-th/0203211}{{\tt hep-th/0203211}}.

\bibitem{Sen2}
A.~Sen, ``Tachyon matter,'' {\em JHEP} {\bf 07} (2002) 065,
\href{http://www.arXiv.org/abs/hep-th/0203265}{{\tt hep-th/0203265}}.

\bibitem{AdSDbrane1}
P.~Lee, H.~Ooguri, and J.-W. Park, ``Boundary states for {$AdS_2$} branes in
  {$AdS_3$},'' {\em Nucl. Phys.} {\bf B632} (2002) 283--302,
\href{http://www.arXiv.org/abs/hep-th/0112188}{{\tt hep-th/0112188}}.

\bibitem{AdSDbrane2}
B.~Ponsot, V.~Schomerus, and J.~Teschner, ``Branes in the {Euclidean}
  {$AdS_3$},'' {\em JHEP} {\bf 02} (2002) 016,
\href{http://www.arXiv.org/abs/hep-th/0112198}{{\tt hep-th/0112198}}.

\bibitem{Y}
K.~P. Yogendran, ``D-branes in 2d lorentzian black hole,'' {\em JHEP} {\bf 01}
  (2005) 036,
\href{http://www.arXiv.org/abs/hep-th/0408114}{{\tt hep-th/0408114}}.

\bibitem{RS}
S.~Ribault and V.~Schomerus, ``Branes in the 2-d black hole,'' {\em JHEP} {\bf
  02} (2004) 019,
\href{http://www.arXiv.org/abs/hep-th/0310024}{{\tt hep-th/0310024}}.

\end{thebibliography}
\end{document}